\documentclass[twocolumn, a4paper, 10pt, showpacs, prb, reprint, aps,superscriptaddress, amsmath, amssymb, amsfonts, floatfix]{revtex4-1}

\usepackage{graphicx, array}
\usepackage{rotating}
\usepackage[caption=false]{subfig}
\usepackage{xcolor}

\begin{document}

\title{Plasmon Spectrum of Single Layer Antimonene}
\author{Guus Slotman}
\affiliation{Institute for Molecules and Materials, Radboud University, Heyendaalseweg 135, 6525AJ Nijmegen, The Netherlands}
\author{Alexander Rudenko}
\affiliation{School of Physics and Technology, Wuhan University, Wuhan 430072, China}
\affiliation{Institute for Molecules and Materials, Radboud University, Heyendaalseweg 135, 6525AJ Nijmegen, The Netherlands}
\author{Edo van Veen}
\affiliation{Institute for Molecules and Materials, Radboud University, Heyendaalseweg 135, 6525AJ Nijmegen, The Netherlands}
\author{Mikhail I. Katsnelson}
\affiliation{Institute for Molecules and Materials, Radboud University, Heyendaalseweg 135, 6525AJ Nijmegen, The Netherlands}
\author{Rafael Rold\'an}
\affiliation{Materials Science Factory. Instituto de Ciencia de Materiales de Madrid (ICMM), Consejo Superior de Investigaciones Cient\'{i}ficas (CSIC), Cantoblanco E28049 Madrid, Spain}
\author{Shengjun Yuan}
\email{s.yuan@whu.edu.cn}
\affiliation{School of Physics and Technology, Wuhan University, Wuhan 430072, China}
\affiliation{Institute for Molecules and Materials, Radboud University, Heyendaalseweg 135, 6525AJ Nijmegen, The Netherlands}

\date{\today}

\begin{abstract}

The collective excitation spectrum of two-dimensional (2D) antimonene is calculated beyond the low energy continuum approximation. The dynamical polarizability is computed using a 6-orbitals tight-binding model that properly accounts for the band structure of antimonene in a broad energy range. Electron-electron interaction is considered within the random phase approximation. The obtained spectrum is rich, containing the standard intra-band 2D plasmon and a set of single inter-band modes. We find that spin-orbit interaction plays a fundamental role in the reconstruction of the excitation spectrum, with the emergence of novel inter-band branches in the continuum that interact with the plasmon. 
\end{abstract}

\maketitle

\section{Introduction}

Antimonene, a single layer of Sb atoms arranged in a buckled honeycomb lattice,\cite{Ares_AM_2017} has been recently fabricated by different methods, from mechanical\cite{Ares_AM_2016} and chemical exfoliation,\cite{Gibaja_ACIE_2016} to epitaxial growth.\cite{Ji_NC_2016,Lei_JAP_2016,Wu_AM_2017} As phosphorene, it is a monoelemental atomically thin crystal of group-VA, with the advantage of being highly stable in ambient conditions. Single layer Sb is an indirect gap semiconductor, with a strong spin-orbit coupling (SOC).  Compared with standard 3D semiconductors such as gallium arsenide or silicon, 2D materials present many special  features like  quantum confinement in the direction perpendicular to the layer, tuneability of the bandgap, or easy integrability in optoelectronic structures, what make them good candidates for nanophotonics.\cite{Xia_NR_2014} 

A huge attention has been paid to the study of plasmons\cite{Grigorenko_NP_2012} and polaritons\cite{Low_NM_2017} of 2D materials like graphene,\cite{Wunsch_NJP_2006,Hwang_PRB_2007,Polini_PRB_2008,Stauber_JPCM_2014}  black phosphorus,\cite{Low_PRL_2014,Fengping_PRB_2015,Ghosh_PRB_2017} or hexagonal boron nitride (h-BN).\cite{Dai_S_2014} Antimonene, like other families of 2D materials,\cite{Roldan_CSR_2017}  presents optical and electronic properties that can be manipulated and tuned by controlled thickness (number of layers), applied strain, external electric fields, or chemical functionalization.  In particular, while a single layer of antimonene is a trivial semiconductor, increasing the number of layers can lead to a transition to a topological semimetal, including the appearance of quantum spin Hall phases. Furthermore, the strong SOC presents in this material combined with its high flexibility makes that strain engineering can be used to drive a transition from a trivial indirect gap semiconductor to a 2D topological insulator.\cite{Zhang_AC_2015,Ares_AM_2017} This may open the door to use antimonene for nanoscale transistors with high on/off ratio, nanomechanical sensors, or optoelectronics devices operating in a broad range of the spectrum. 

 In this paper we present a systematic study of the excitation spectrum of single layer antimonene. The band structure is obtained from a tight-binding (TB) Hamiltonian that considers the 3-$p$ orbitals of each of the two Sb atoms of the unit cell.\cite{Rudenko_PRB_2017} The dielectric function is obtained within the random phase approximation (RPA). Due to the large number of bands considered in the calculation, the obtained excitation spectrum is rich, with several plasmon and inter-band modes. Consideration of SOC is shown to be essential to capture the correct low energy excitations, due to strong reconstruction of the band structure.  
 
 \begin{figure}[ht!]
    \includegraphics[width=0.98\linewidth]{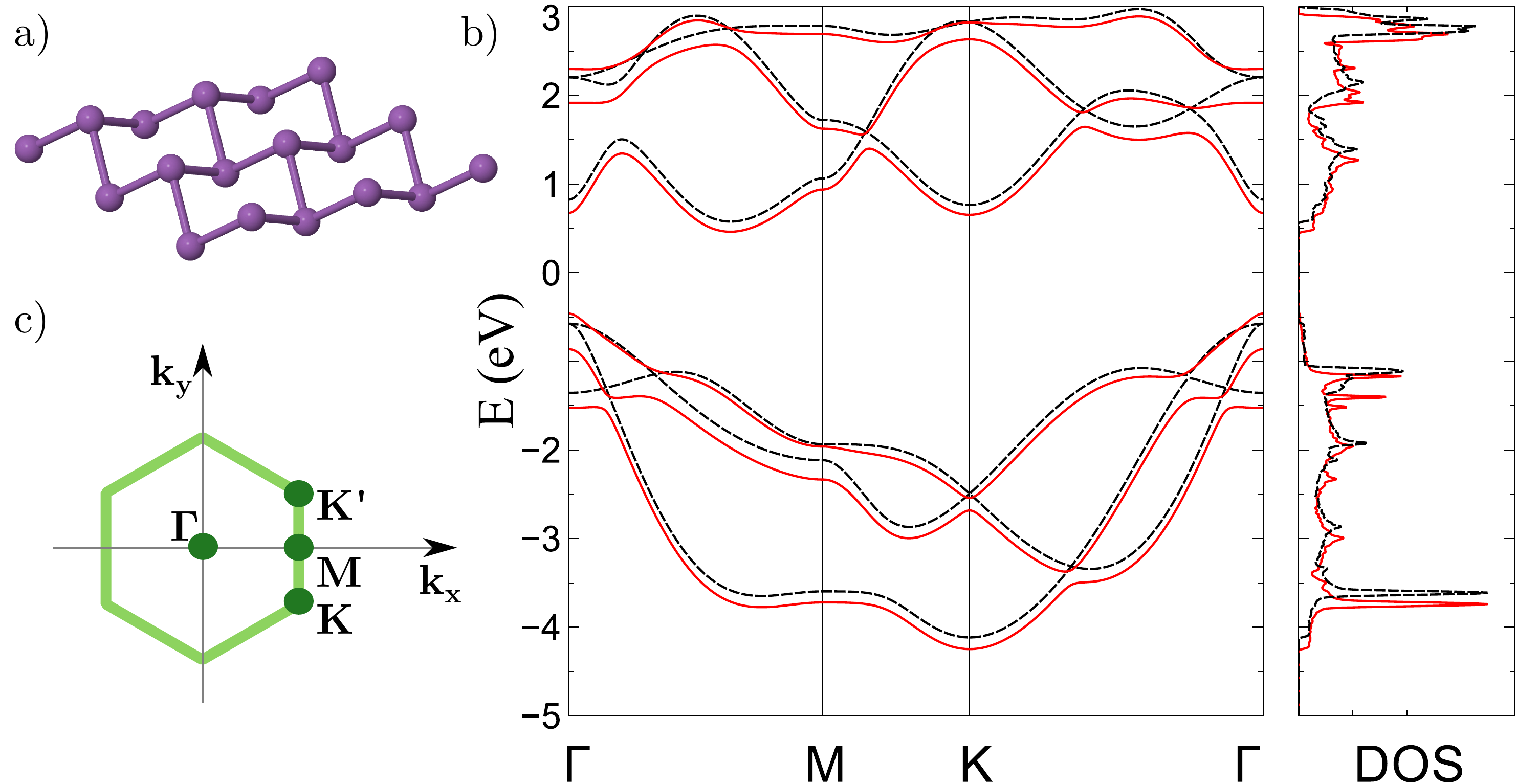}    
    \caption{(a) The buckled atomic structure and (c) Brillouin zone of SL-Sb. (b) The band structure of SL-Sb as obtained in Ref.~\cite{Rudenko_PRB_2017} using their TB-Hamiltonian with (solid red lines) and without (dashed black lines) spin-orbit coupling . }
\label{fig:band}
\end{figure}

\begin{figure*}[ht!]
    \includegraphics[width=0.99\linewidth]{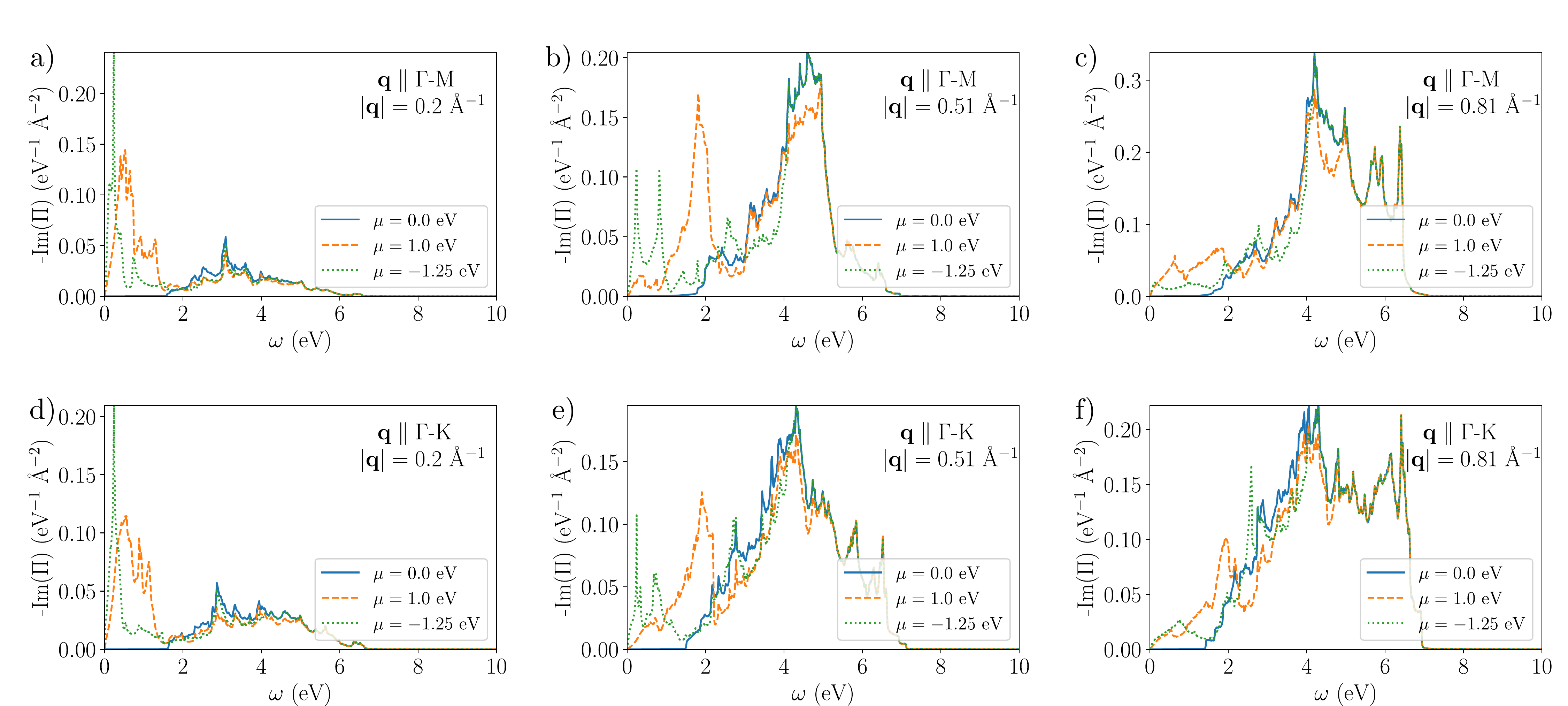}
    \caption{Imaginary part of the polarization function $\Pi({\bf q}, \omega)$ for different values of wave-vector $\bf q$ and chemical potential $\mu$. In the top images $\bf q$ lies on the $\Gamma$-M line in the BZ, while in the bottom images $\bf q$ lies on the $\Gamma$-K line.}
\label{fig:dp_imag_lineplots}
\end{figure*}

\section{Method}
Single layer antimonene has a buckled honeycomb structure with lattice parameter $a=4.12$~\AA , where the two sublattices are vertically displaced by $b=1.65$~\AA. The band structure, including SOC, can be well described by the tight-binding model developed in Ref. \onlinecite{Rudenko_PRB_2017}. The TB model Hamiltonian is given by:
\begin{equation}
{\cal H} =  \sum_{m,n} \sum_{i,j} t_{ij}^{mn} c_{im}^\dagger c_{jn}+{\cal H}_{SO},
\label{eq:sb_tb-hamil}
\end{equation}
where $t_{ij}^{mn}$ is the hopping parameter between the $m$ and $n$ orbitals at sites $i$ and $j$. The model considers the three $p$ orbitals of Sb, the hopping parameters being obtained from the formalism of maximally-localized Wannier functions. $15$ relevant terms are included, corresponding to hopping between atoms separated up to a distance of $8.24$~\AA. Intra-atomic SOC is considered by the term ${\cal H}_{SO}=\lambda~ {\bf L}\cdot {\bf S}$, where $\bf L$ and $\bf S$ are the total atomic angular momentum operator and the total electronic spin operator, respectively, and $\lambda=0.34$~eV is the intra-atomic SOC constant obtained for Sb. The band structure obtained from this TB model, Fig. \ref{fig:band}, fairly reproduces the spectrum obtained from first principles density functional theory (DFT) methods in an energy range from $[-4,+3]$~eV.  We notice the important role played by SOC, that reconstructs the band structure leading to an overall reduction of the indirect gap in $\sim 0.3$~eV, as well as to a number of avoided crossings between some bands. Those will play an important role when studying the single particle excitation spectrum, as we will see below.

The TB-Hamiltonian (\ref{eq:sb_tb-hamil}) is used to numerically calculate the dynamical polarization $\Pi (\mathbf{q},\omega)$ and the dielectric function $\epsilon(\mathbf{q},\omega)$ using the Lindhard function\cite{Giuliani_Book_2005}

\begin{widetext}
\begin{equation}
\Pi (\mathbf{q},\omega) = - \frac{g_s}{(2\pi)^2} \int_{BZ} d^2 \mathbf{k} \sum_{l,l'} \frac{n_F (E_{\mathbf{k} l}) - n_F (E_{\mathbf{k'} l'})}{E_{\mathbf{k} l} -E_{\mathbf{k'} l'} + \omega + i\delta} \left| \left< \mathbf{k'} l' | e^{i \mathbf{q} \cdot r} | \mathbf{k} l \right> \right|^2 ,
\label{eq:lindhard}
\end{equation}
\end{widetext}
where $ | \mathbf{k} l \rangle $ and $E_{\mathbf{k} l}$ are eigenstates and eigenvalues of the Hamiltonian (\ref{eq:sb_tb-hamil}), respectively, with $l$ being the band index, $\mathbf{k'} = \mathbf{k} + \mathbf{q}$, $n_F (E) = \frac{1}{e^{\beta(E-\mu)}+1}$ is the Fermi-Dirac distribution, $\delta\rightarrow 0^+$ and we have taken $\hbar=1$. The integral is taken over the Brillouin zone (BZ), and the sum  is calculated over all bands in the TB Hamiltonian. From the dynamical polarization one can obtain the dielectric function within the RPA as:
\begin{equation}
\epsilon (\mathbf{q},\omega)  = \mathbf{1} - V(q) \Pi (\mathbf{q},\omega),
\label{eq:dielectric}
\end{equation}
with $V(q) = \frac{2 \pi e^2 }{\epsilon_B q}$ the Fourier component of the Coulomb interaction in two dimensions, with  $\epsilon_B$ being the background dielectric constant. We use $\epsilon_B = 3.9$ to represent a dielectric substrate.

In order to explore the long wavelength limit of the dielectric function of single layer Sb, we make use of the following relation with the imaginary part of the two-dimensional optical conductivity \cite{Seybold_Book_2005}

\begin{equation}\label{Eq:q=0}
	\text{Re}~ \epsilon_{\alpha \alpha}(\mathbf{q} = 0, \omega) = 1 - \frac{\text{Im}~ \sigma_{\alpha \alpha}(\omega)}{\omega d \epsilon_B},
\end{equation}
where $d$ is the height of  single layer Sb, which can be approximated as the interlayer distance in its bulk structure ($d = 0.36$ nm), and $\text{Im} \sigma_{\alpha \alpha}(\omega)$ is obtained by using the Kramers–-Kronig relations from the optical conductivity along the $\alpha$-direction

\begin{eqnarray}
	\text{Re}(\sigma_{\alpha \alpha}(\omega)) = -\frac{g_S}{\Omega \omega} \int_{BZ}  \text{ Im} \left[ \sum_{l,l'} |\langle \mathbf{k} l | J_{\alpha} | \mathbf{k} l' \rangle| ^2 \right. \nonumber \\
	\left. \times \frac{f(E_{ \mathbf{k}l}- \mu) - f(E_{ \mathbf{k}l'}- \mu)}{E_{ \mathbf{k}l} - E_{ \mathbf{k}l'} + \omega + i \delta} \vphantom{\sum_{l,l'}} \right] d^2 \mathbf{k} ,
\end{eqnarray}

where $\Omega$ is the unit cell surface, and $J_{ \alpha}$ is the current operator in the $\alpha$-direction

\begin{equation}
	J_{ \alpha} = - \frac{i e}{\hbar} \sum_{i, j}  t_{ij}  (\mathbf{r}_j - \mathbf{r}_i)_{\alpha} c_{\mathbf{k}i}^{\dagger} c_{\mathbf{k}j}^{\phantom{\dagger}}.
\end{equation}

\begin{figure}[t!]
    \includegraphics[width=0.99\linewidth]{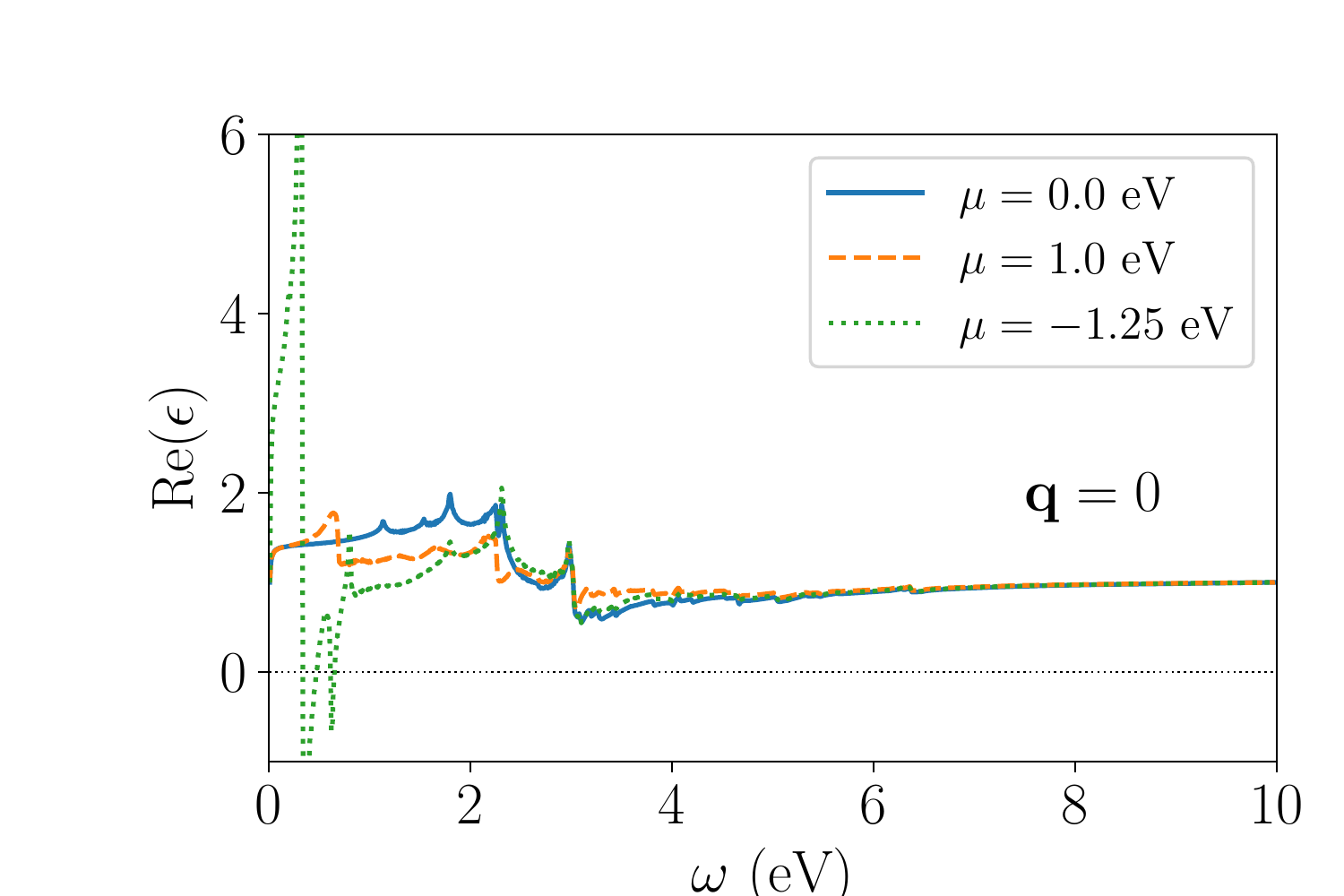}
    \caption{{Real part of the dielectric function $\epsilon (0,\omega)$ for different chemical potential $\mu$, calculated using Eq.~(\ref{Eq:q=0}). }} 
\label{fig:ep_real_q0}
\end{figure}

For isotropic materials such as antimonene, the subscript $\alpha $ in the dielectric function can be ignored as  $ \epsilon$ is also isotropic and $ \epsilon_{x x}=\epsilon_{y y}$.
The dielectric function of antimonene in the long wavelength limit is given, for different values of the chemical potential, in Fig. \ref{fig:ep_real_q0}. For the case of $\mu = -1.25 $~eV, the zeros of the dielectric function correspond to the inter-band modes that will be discussed in Sec. \ref{Sec:Discussion}.

\section{Results}

\subsection{Dynamical Polarization}
\label{sec:dp}

\begin{figure*}[t!]
    \includegraphics[width=0.99\linewidth]{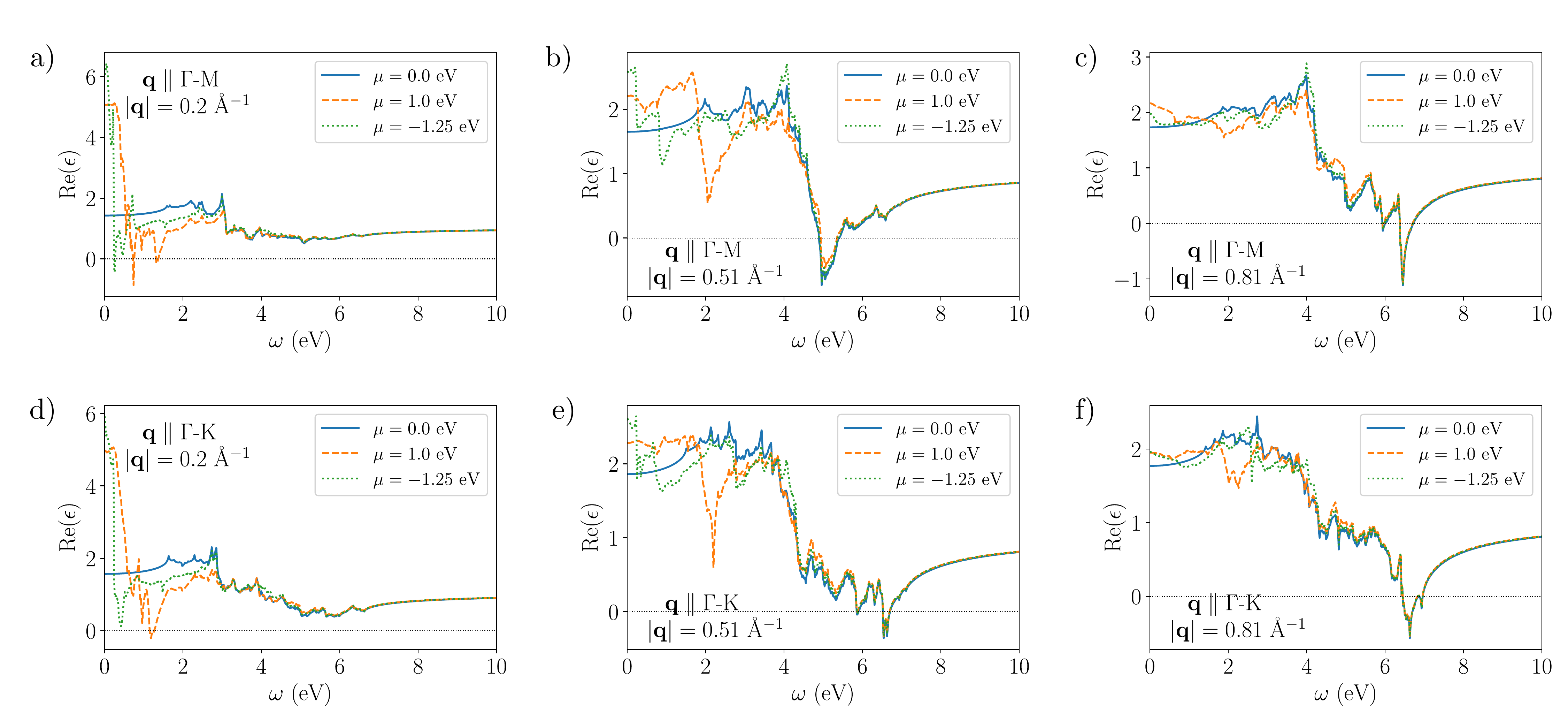}
    \caption{Real part of the dielectric function $\epsilon (\mathbf{q},\omega)$ for different wave-vector $\bf q$ and chemical potential $\mu$. In the top images $\bf q$ lies along the $\Gamma$-M direction of the BZ, while in the bottom images $\bf q$ lies along $\Gamma$-K. For a given $\bf q$ and $\mu$, the energy of the collective plasmon mode is given by the zeros of the dielectric function.}
\label{fig:ep_real_lineplots}
\end{figure*}

The particle-hole excitation spectrum of the 2D electron gas in single layer antimonene is defined, as usual, as the region of the energy-momentum
space that is available for electron-hole excitations. For non-interacting
electrons, it is defined as the region where $\mathrm{Im}~\Pi (\mathbf{q}%
,\omega )$, as given by Eq. (\ref{eq:lindhard}), is
non-zero.\cite{Giuliani_Book_2005} Figure~\ref{fig:dp_imag_lineplots} shows the dynamical polarization function for different directions of $\bf q$ (along $\Gamma$-M and $\Gamma$-K) and different chemical potentials $\mu$, at $T=300$~K. 
Due to the high number of bands taken into account in the calculation, the spectrum is rich in features with contributions from many interband excitations. Although similar, the dynamical polarization for both directions of the wave-vector features some different excitations. We observe that modifying the chemical potential in Eq.~(\ref{eq:lindhard}) leads to the inclusion, or exclusion, of additional interband electron-hole excitations, resulting in extra low-energy peaks in the dynamical polarization. For a given wave-vector $\bf q$, we can see that the spectra for different values of $\mu$ present similar structure at high energies, while they are notably different at low energies. The high energy sector of the polarization function $\Pi({\bf q},\omega)$ stays mostly undisturbed because it is built by inter-band transitions involving the lower (higher) energy bands of the hole (electron) sectors. For undoped antimonene ($\mu=0$), only inter-band transitions are allowed, and therefore $\mathrm{Im}~\Pi (\mathbf{q},\omega )=0$ for frequencies below the bandgap $\omega<\Delta$.\footnote{For a more complete visualisation of the excitation spectrum, in Fig. \ref{fig:dp_imag_contour} of the Appendix we show a density plot of $\mathrm{Im}~\Pi (\mathbf{q},\omega )$ in the $\omega-{\bf q}$ plane.}

When we consider electron or hole doping, we obtain a finite contribution to the spectral weigh at low energies, due to activation of intra-band electron-hole transitions. These are the processes that dominate the spectrum at long wavelengths, as it is seen for the small wave-vector sector of Figs.  \ref{fig:dp_imag_lineplots} (a) and (d) (see also Figs. \ref{fig:dp_imag_contour} (b),(c) of the Appendix).

\subsection{Dielectric Function and Collective Modes}
\label{sec:ep}

\begin{figure*}[ht!]
    \includegraphics[width=0.32\linewidth]{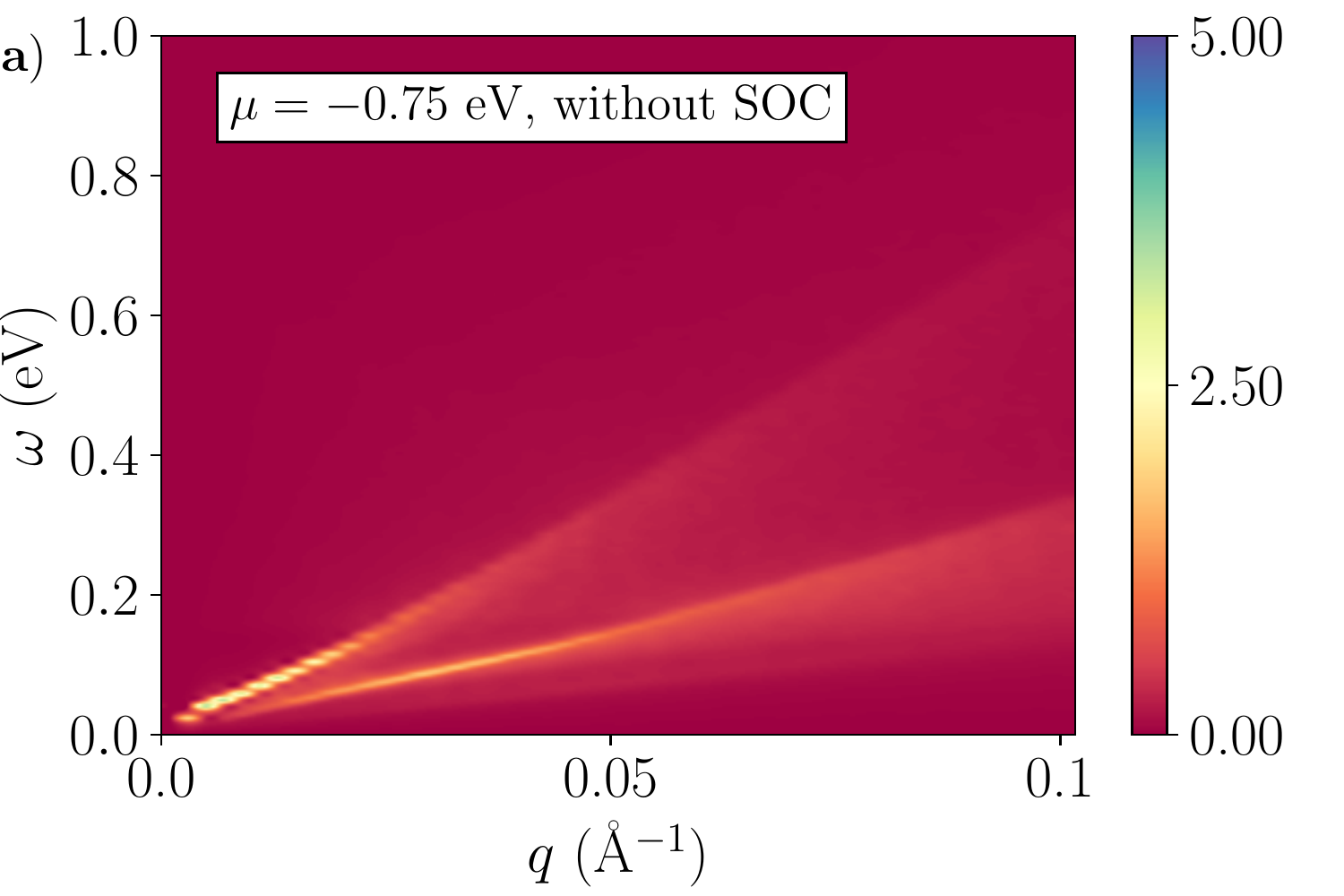}
    \includegraphics[width=0.32\linewidth]{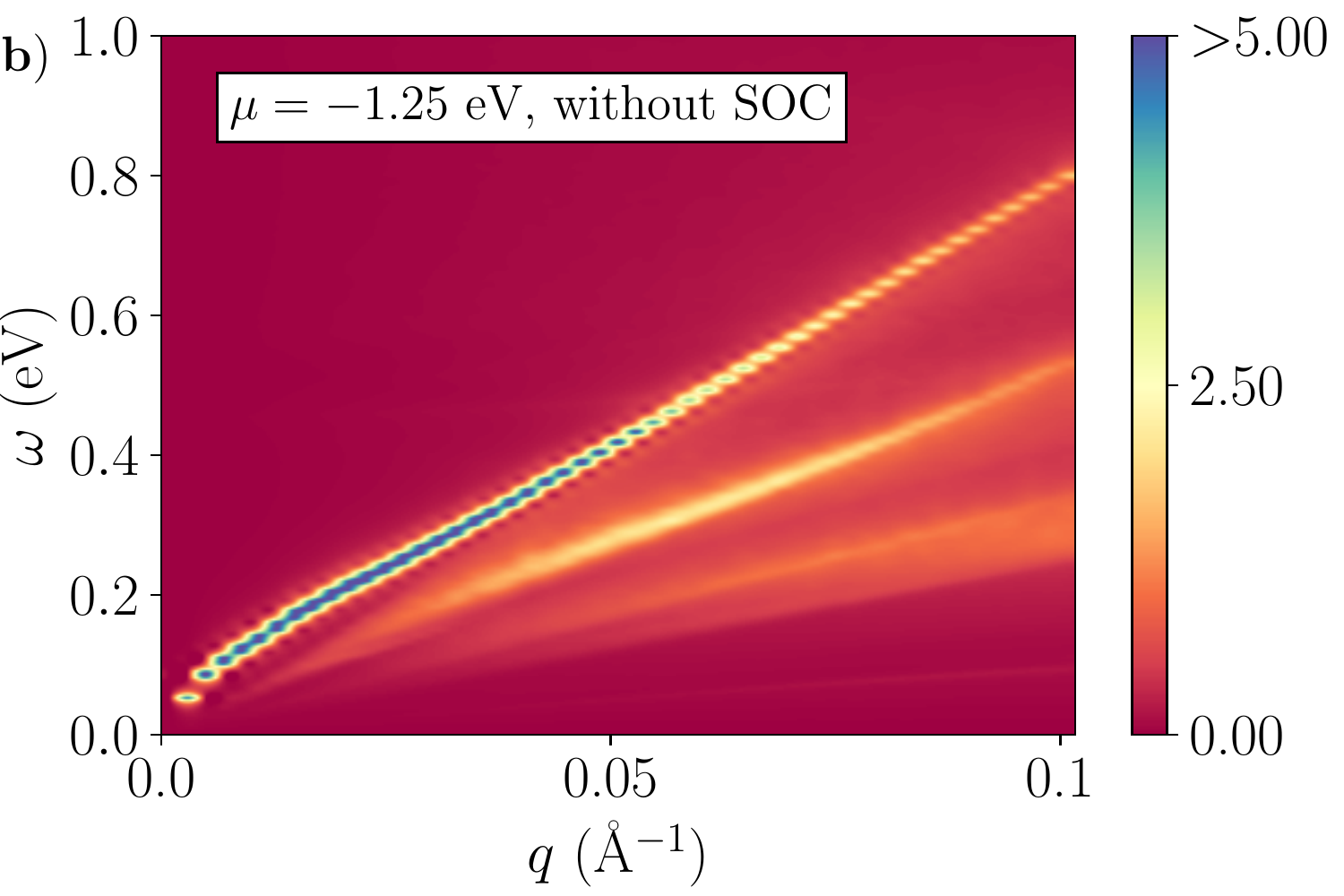}
    \includegraphics[width=0.32\linewidth]{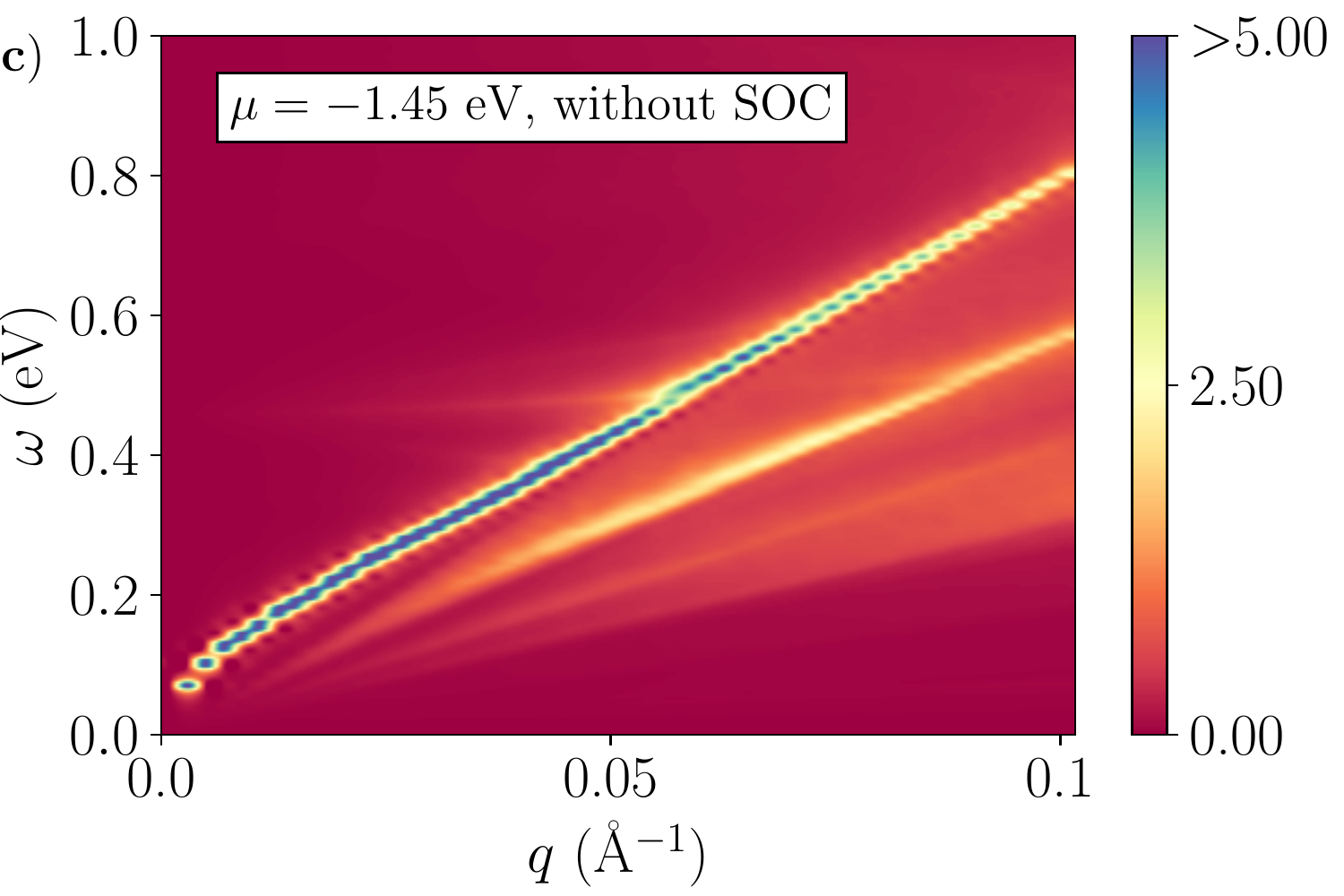}
    \includegraphics[width=0.32\linewidth]{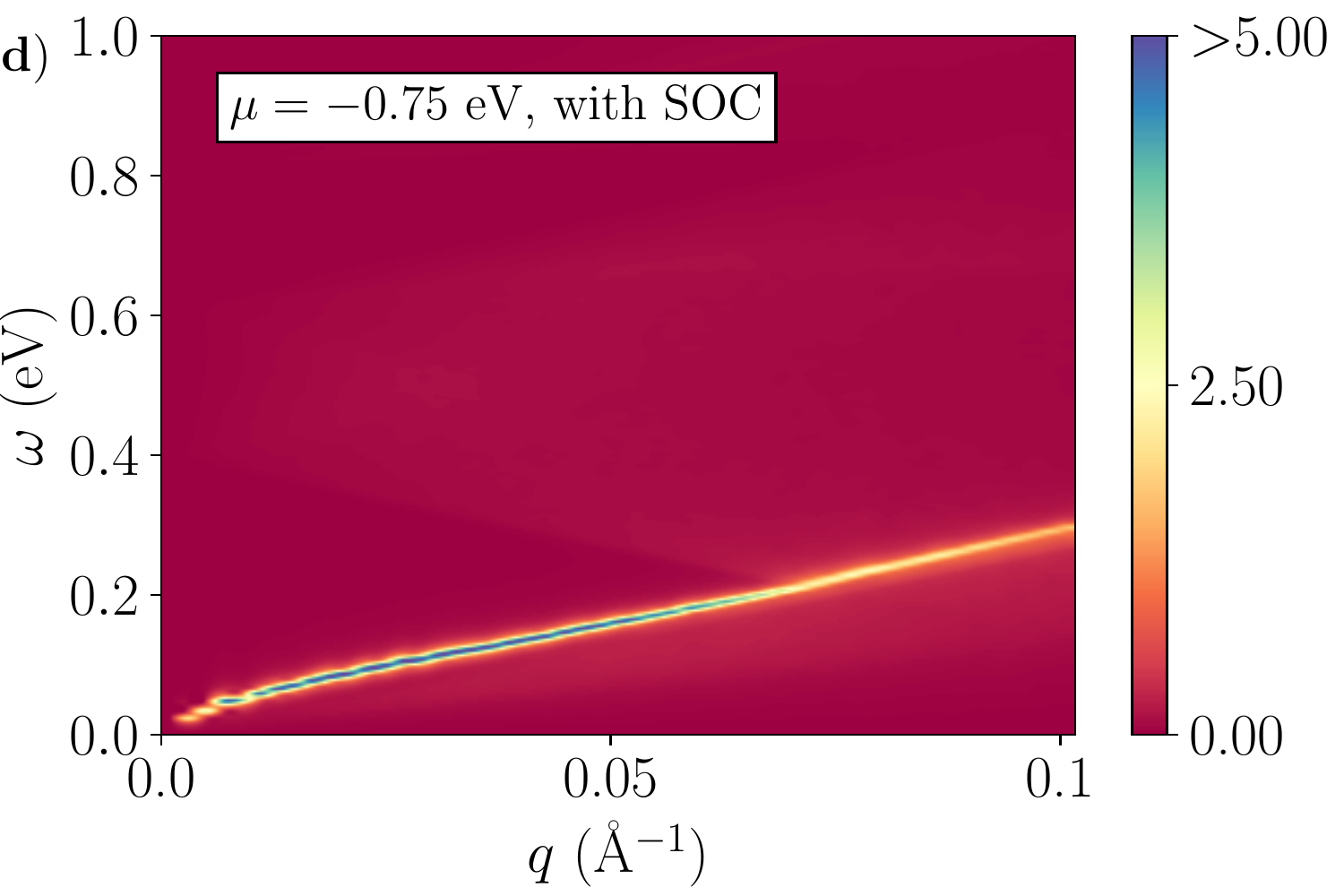}
    \includegraphics[width=0.32\linewidth]{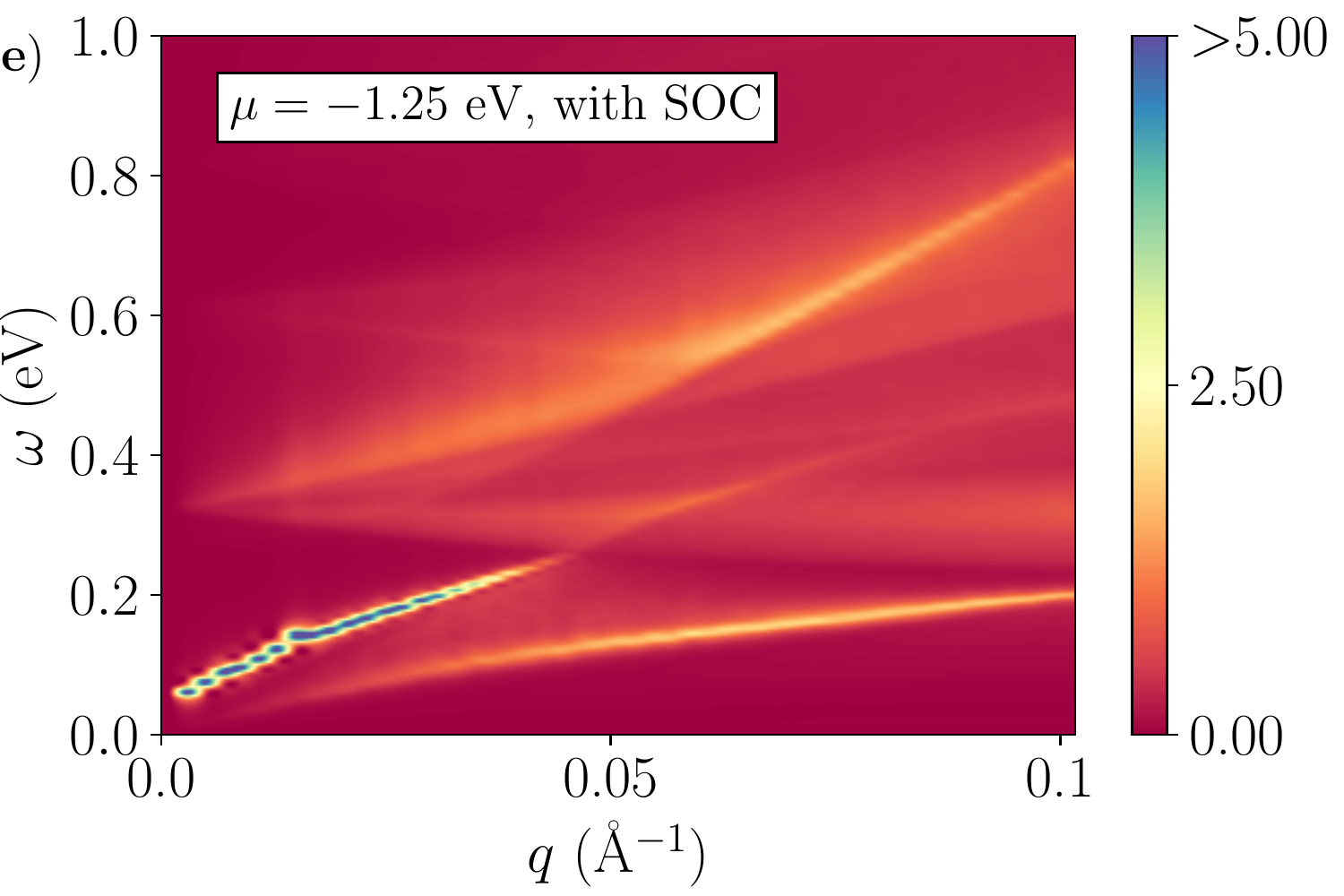}
    \includegraphics[width=0.32\linewidth]{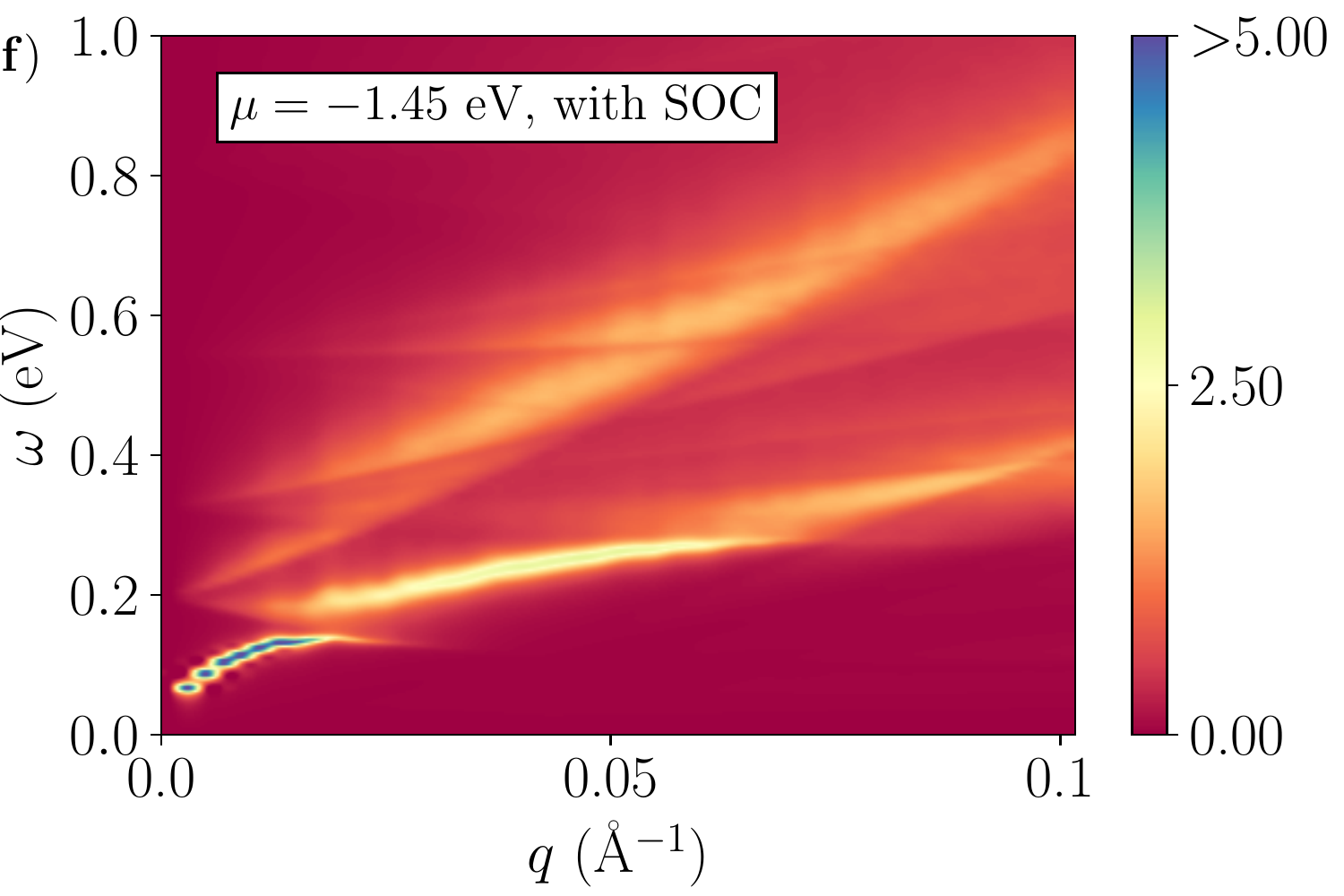}
    \caption{The loss function  $-{\rm Im}(1/\epsilon)$ in the low energy and wave-vector region of the spectrum. 
    Panels in the first row correspond to $-{\rm Im}(1/\epsilon)$ without SOC, while the second row is with SOC. For each image the loss function is calculated for 50 q-points, with fitted points in between. Due to this some of the figures have a chopped appearance. }  
\label{fig:loss_contour_small_q}
\end{figure*}

The next step is to consider electron-electron interaction in the spectrum. Within the RPA, the dielectric function of the system $\epsilon (\mathbf{q},\omega)$ is calculated from (\ref{eq:dielectric}), and the corresponding results are shown in Fig.~\ref{fig:ep_real_lineplots}. As usual, the zeros of the dielectric function will define the existence and dispersion of collective plasmon modes. There are two distinct regions were ${\rm Re}{\left[\epsilon (\mathbf{q},\omega)\right]}=0$, indicating the existence of several modes. In particular, there is a high energy plasmon mode at around $\omega\sim7$~eV.  The exact location of these modes depend on both, the chemical potential $\mu$ and the wave-vector $\mathbf{q}$ in the BZ. We must notice that, although this mode corresponds to a zero of the dielectric function, 
as it can be seen in Fig. \ref{fig:ep_real_lineplots}, it is a mode whose dispersion lies  
\textit{inside} the continuum of electron-hole excitations: $-\mathrm{Im}%
~\Pi(\mathbf{q},\omega_{pl})>0$ at the plasmon energy $\omega_{pl}$, what implies that the mode is damped, continuously decaying into electron-hole pairs.  Nevertheless, it is a
well defined mode that could be measured experimentally using electron energy loss spectroscopy (EELS),\cite{Egerton_Book_2011} technique that has been successfully applied to study the plasmon spectrum of graphene\cite{Eberlein_PRB_2008} and other 2D materials.\cite{Politano_MSSP_2017}

As discussed in the previous section, changing the chemical potential only has noticeable effect for energy values smaller than $\sim 4$~eV.\footnote{See also Fig.~\ref{fig:ep_real_contour} of Appendix for a density plot of ${\rm Re}~\varepsilon({\bf q}, \omega)$ in the $\omega-{\bf q}$ plane.} In the presence of finite doping $\mu$ and for small wave-vectors, the standard low-energy plasmon mode is recovered. The main features of the low energy spectrum will be discussed in Sec. \ref{Sec:Discussion}.

\section{Discussion}\label{Sec:Discussion}

\begin{figure}[ht!]
    \includegraphics[width=0.8\linewidth]{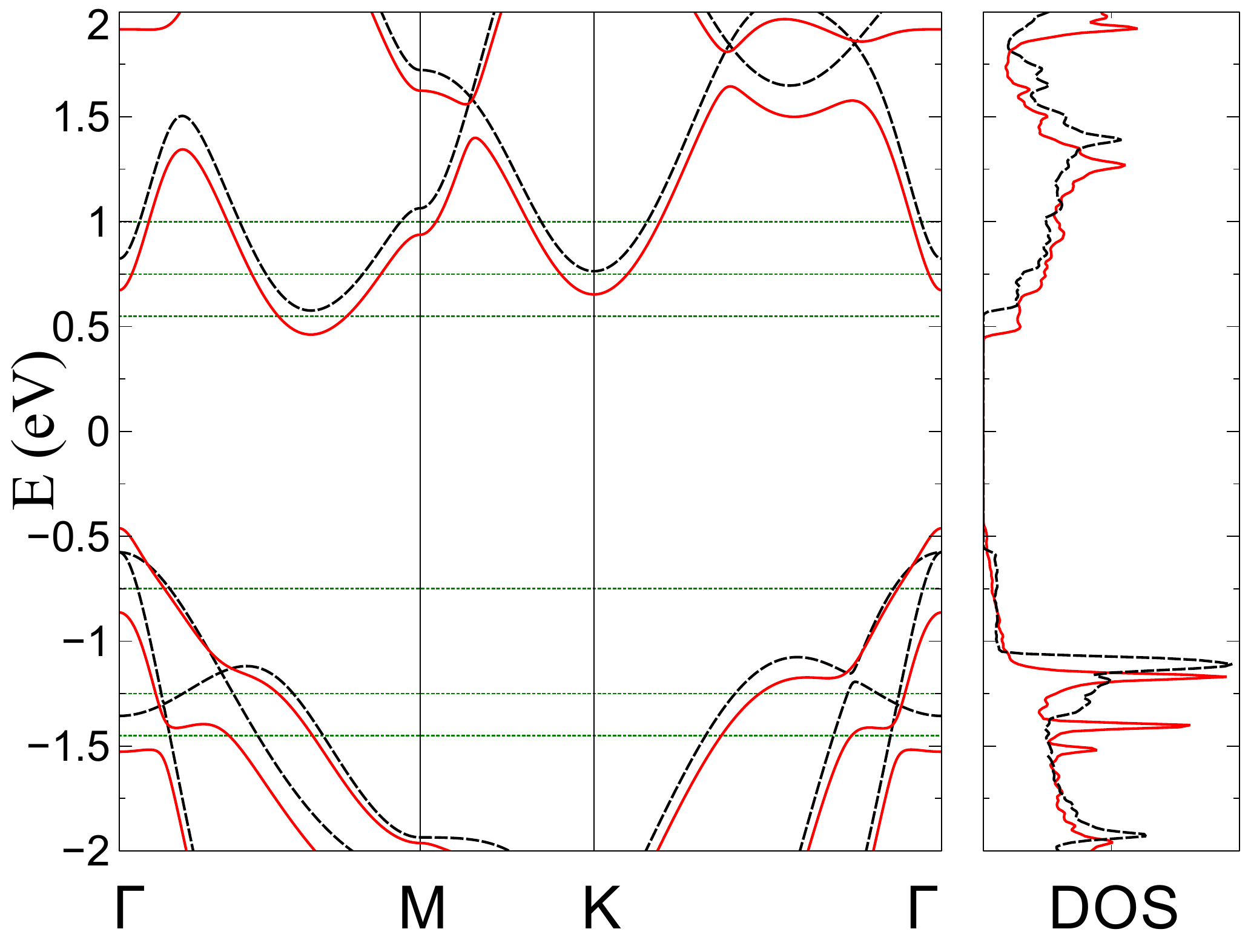}    
    \caption{Comparison of the band structure and density of states of single layer antimonene with (solid red) and without (dashed black) spin-orbit coupling. Green dotted lines show the values of the chemical potential $\mu$ used in the calculations of the loss function in Fig. \ref{fig:loss_contour_small_q}.}
\label{Fig:Bands_LowEnergy}
\end{figure}

\begin{figure*}[ht!]
    \includegraphics[width=0.99\linewidth]{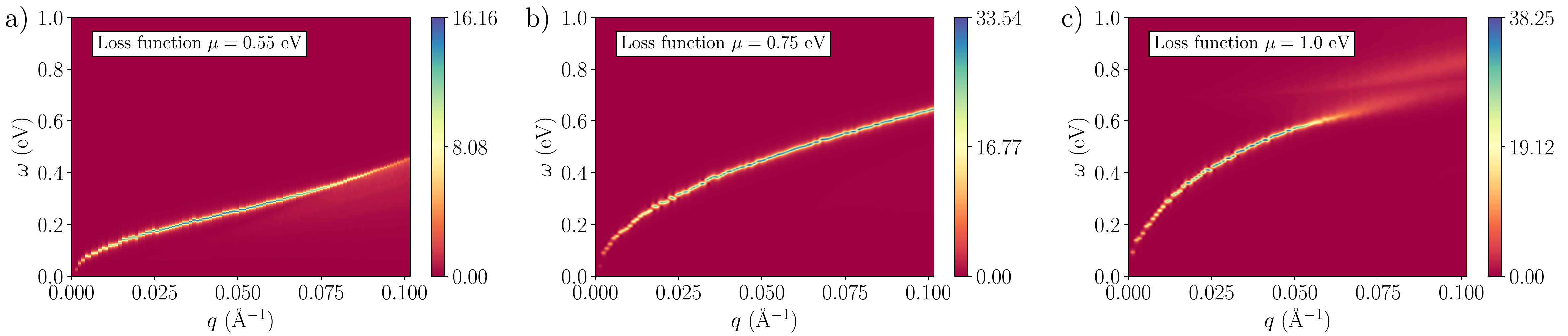}

    \caption{The loss function  $-{\rm Im}(1/\epsilon)$ in the low energy and wave-vector region of the spectrum for electron doping. For each image the loss function is calculated for 50 q-points, with fitted points in between. Due to this some of the figures have a chopped appearance.}  
\label{fig:loss_contour_small_q_electron}
\end{figure*}

In this section we focus on the low energy region of the excitation spectrum, which is the most relevant for future experimental probes and possible applications. Fig.~\ref{fig:loss_contour_small_q} shows the loss function for hole doped antimonene with $\omega \leq 1 eV $and $q  \leq 0.1$~\AA$^{-1}$ for different values of $\mu$, including or not SOC in the calculation. Each contour plot is extrapolated from 50 $q$ points. The first thing we notice is the existence of an intra-band plasmon with a dispersion $\propto \sqrt{q}$. This is expected for any 2D electron gas, and constitutes the standard collective oscillation of carriers due to long range Coulomb interaction.\cite{Stern_PRL_67} The dispersion of this mode, at low energies and for $q\rightarrow 0$, is
\begin{equation}\label{Eq:Plasmon}
\omega_{pl}(q)\approx\sqrt{\frac{2e^2\mu}{\varepsilon_b}q+\frac{3}{4}v_F^2q^2}
\end{equation}
where $v_F=\partial \epsilon/\partial k|_{k=k_F}$ is the Fermi velocity and $k_F=\sqrt{4\pi n/g}$ is the Fermi wave-vector, in terms of the carrier density $n$ and the degeneracy factor $g=g_sg_v$, where $g_s=2$ is the spin degeneracy and $g_v=1$ is the valley degeneracy for carriers in a single hole pocket around the $\Gamma$ point of the BZ.    

By looking at the loss spectrum for $\mu=-0.75$~eV in Fig. \ref{fig:loss_contour_small_q}(d), we can corroborate the existence of a mode whose dispersion responds to Eq.~(\ref{Eq:Plasmon}). To analyze the effect of SOC, let us compare Fig. \ref{fig:loss_contour_small_q}(a) and (d).  If SOC is neglected, there are two bands (labelled here as $i=1,2$) that are occupied for $\mu=-0.75$~eV. See Fig. \ref{Fig:Bands_LowEnergy} for a clear visualization of the band filling at given $\mu$, for the cases with (solid red) and without (dashed black) SOC. Notice that, if SOC is not considered, these bands are degenerate exactly at the $\Gamma$ point. The loss function for this case, Fig. \ref{fig:loss_contour_small_q}(a), shows two well defined branches. The branch with larger slope is the plasmon, while the other structure is originated from the interaction between the two intra-band single-particle continua associated to bands 1 and 2.
The equation for the plasmon in this case has the form{\cite{sarma_prb_1984}}
\begin{eqnarray}\label{Eq:TwoPockets}
[1-V(q)\Pi^0_{1}({\bf q},\omega)][1-V(q)\Pi^0_{2}({\bf q},\omega)]&&\nonumber\\
-V(q)^2\Pi^0_1({\bf q},\omega)\Pi^0_{2}({\bf q},\omega)&=&0
\end{eqnarray}
where the polarization function of the carriers of band $i=1,2$ can be approximated, in the dynamical long wavelength limit, as\cite{Giuliani_Book_2005}
\begin{equation}
\Pi^0_{i}({\bf q},\omega)\approx \frac{n_i}{m_i}\frac{q^2}{\omega^2}
\end{equation}
where $n_i$ and $m_i$ are the corresponding density and effective mass of carriers in pocket $i$. The solution of (\ref{Eq:TwoPockets}) gives a plasmon mode with low energy dispersion
\begin{equation}\label{Eq:Plasmon_2pockets}
\omega_{pl}(q)\propto \sqrt{\left(\frac{n_1}{m_1} +\frac{n_2}{m_2} \right )q}.
\end{equation}

 However, the role of SOC has been proven to be very important in antimonene,\cite{Rudenko_PRB_2017} leading  to a reconstruction of the band structure that, in particular, lifts the degeneracy between those two bands at the $\Gamma$ point (see Fig.~\ref{Fig:Bands_LowEnergy}). Therefore, a reliable calculation that includes SOC considers, for the same value of $\mu\approx -0.75$~eV, carriers filling only one subband, with the corresponding reduction (for similar values of the effective masses) in the slope of the dispersion, as it is indeed observed in Fig. \ref{fig:loss_contour_small_q}(d).

Inclusion of SOC is not only necessary to get the correct dispersion relation of the plasmon. The reconstruction of the band structure after consideration of SOC is such, that several avoided crossings that occur lead to emergence of new modes in the spectrum, as it can be seen by comparing Fig. \ref{fig:loss_contour_small_q} (b) and (e), or Fig. \ref{fig:loss_contour_small_q}~(c) and (f). The new modes, seen as bright branches in the  excitation spectrum, are gapped single-particle resonances associated to inter-band transitions. The number of these modes is determined by the possible inter-band transitions from occupied to unoccupied states. Their intensity depend on the DOS at the corresponding energy, which are enhanced for flat bands (saddle points) in the dispersion, associated to the aforementioned avoided crossings in the band structure due to SOC  (see lateral panel of Fig.~\ref{Fig:Bands_LowEnergy} for a comparison of the DOS). Importantly, these inter-band modes interact with the intra-band plasmon, what becomes a source of damping for this mode, effect that is absent when SOC is neglected. Such effect can be easily seen in Figs. \ref{fig:loss_contour_small_q}(e) and (f), where the intra-band $\sim\sqrt{q}$ plasmon interact with inter-subband continuum, with boundaries $E_{ij}\pm v_Fq + q^2/2m$ (assuming equally dispersing parabolic bands), where $E_{ij}$ is the energy separation between bands $i$ and $j$. When the intra-band plasmon hits the inter-subband continuum, the mode is damped, decaying into inter-band electron-hole pairs, leading to avoided crossings in the dispersion relation. This effect is reminiscent of the interaction of graphene plasmons with optical phonons,\cite{Yan_NP_2013} and could be also observed with mid-infrared transmission measurements, or with EELS experiments, as discussed experimentally\cite{Eberlein_PRB_2008,Politano_MSSP_2017} and theoretically\cite{Yuan_PRB_2011,novko2015changing,kupvcic2016effective} for graphene

We finally consider the case of electron doping.  In the low energy sector of the conduction band, SOC only leads to some shifts of the band edges, with no qualitative difference with respect to the spectrum without SOC (see Fig. \ref{Fig:Bands_LowEnergy}). Therefore we present only the excitation spectrum of antimonene for the electron doped regime including SOC. The loss function, Fig. \ref{fig:loss_contour_small_q_electron}, shows the existence of a well defined plasmon branch whose slope increases with doping. This is expected because, for a single occupied band, the velocity of the mode grows with $\mu$, Eq. (\ref{Eq:Plasmon}). Furthermore, as discussed above, the slope also increases as one start filling new pockets of the band structure, as given by Eq. (\ref{Eq:Plasmon_2pockets}).

\section{Conclusion}
In summary, we have calculated the excitation spectrum for single layer antimony, using a 6-orbital tight-binding Hamiltonian that includes SOC. The dielectric function is calculated within the RPA. We obtain a rich spectrum that contains standard plasmons and a set of inter-band modes, originated from the band reconstruction due to SOC. At high energies, the obtained loss functions shows a broad peak at $\sim 6$~eV, associated to the existence of a plasmon, which could be measured by EELS experiments.\cite{Politano_MSSP_2017} At low energies, the interaction between the plasmon and the inter-band single-particle continuum leads to damping of this collective mode.   We find avoided crossing features in the dispersion of the modes that could be observed with mid-infrared transmission measurements.\cite{Yan_NP_2013}

 \acknowledgments 
 
RR acknowledges financial support from the Spanish MINECO through Ram\'on y Cajal program,  Grants No. RYC-2016-20663 and Grant No. FIS2014- 58445-JIN.
M.I.K. acknowledges financial support from the European Research Council Advanced Grant program (contract 338957). 
Yuan acknowledges financial support from Thousand Young Talent Plan (China).
This work as sponsored by NWO Exact and Natural Sciences for the use of supercomputer facilities.

\appendix

\section{}

\begin{figure*}[b]
    \includegraphics[width=0.32\linewidth]{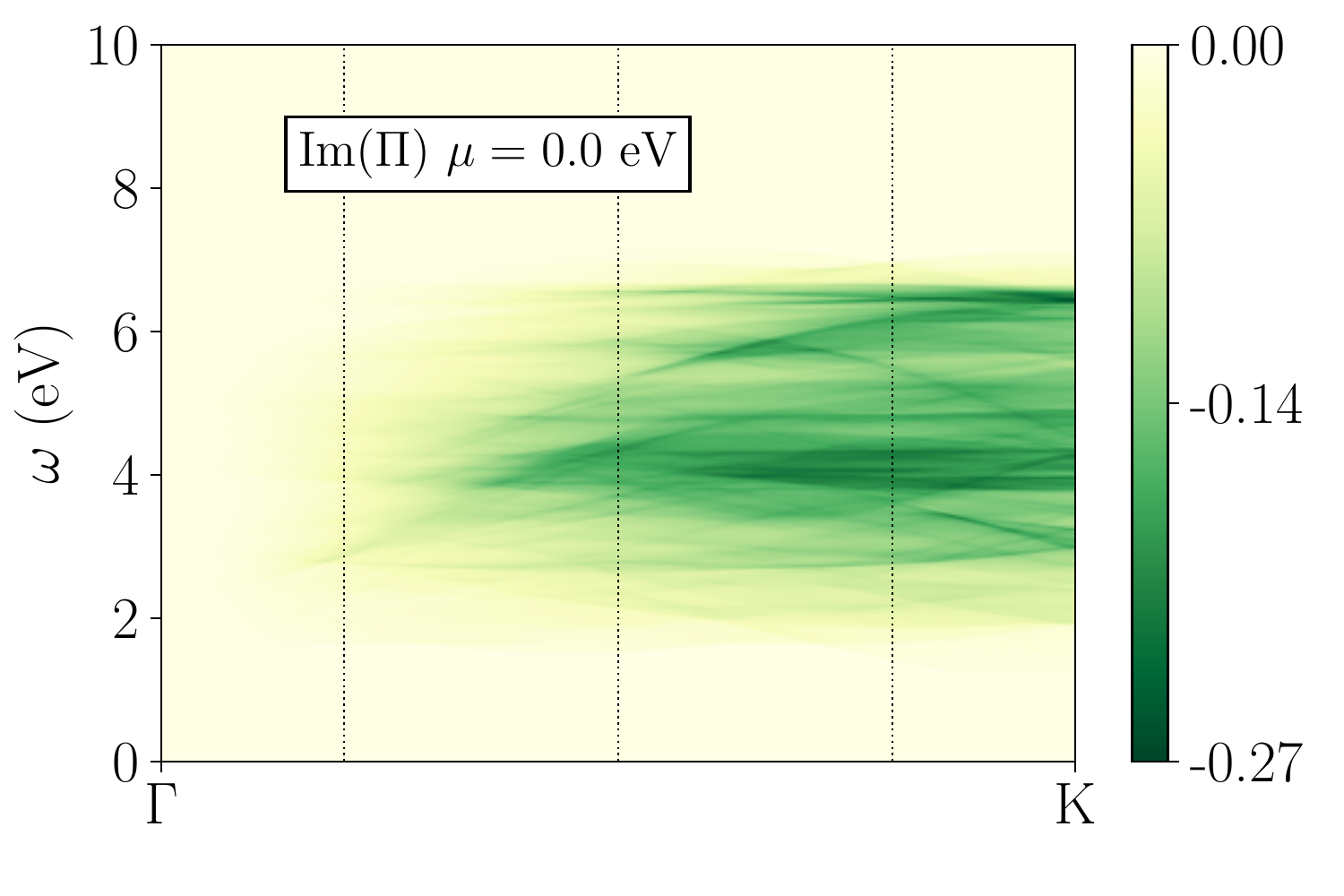}
    \includegraphics[width=0.32\linewidth]{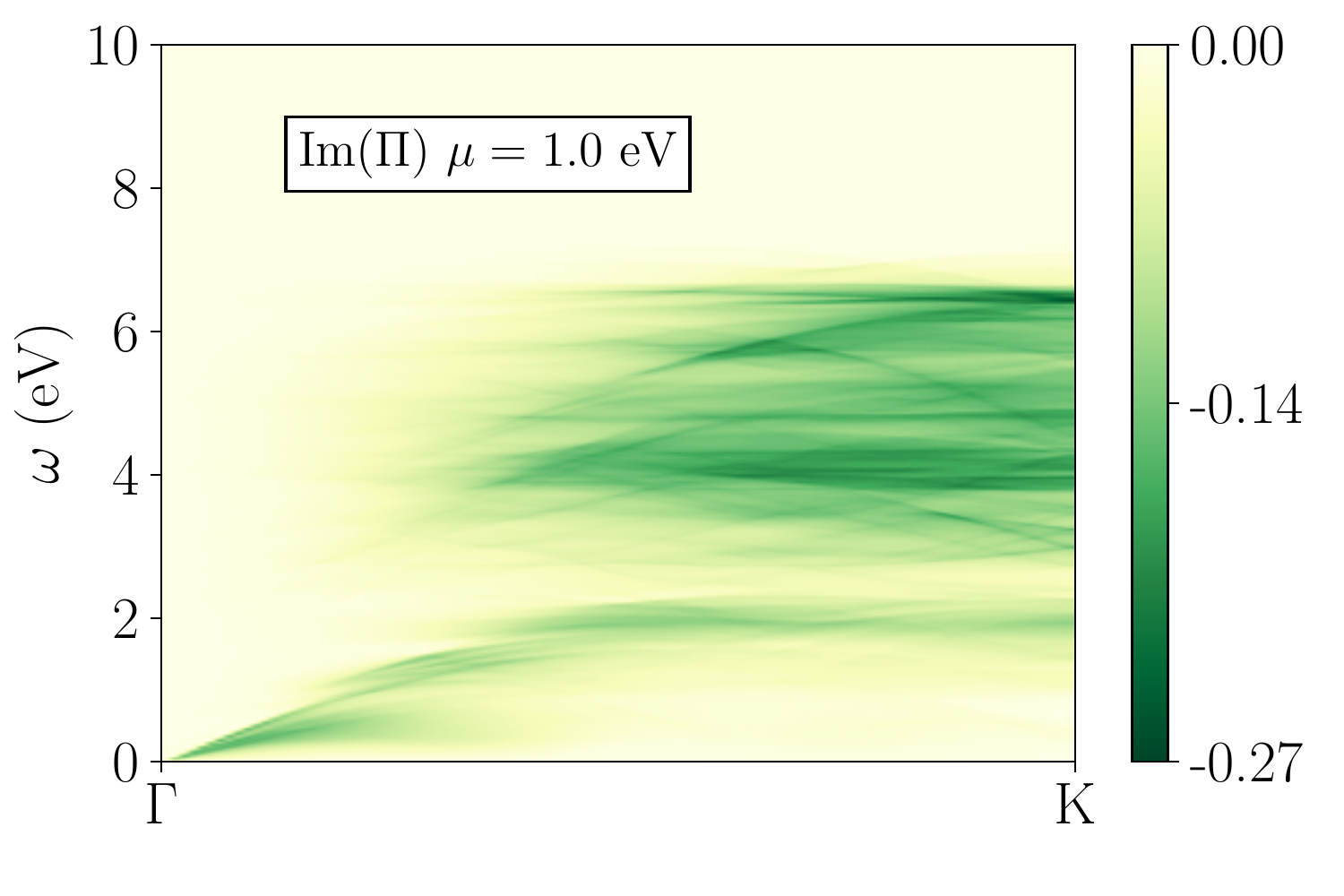}
    \includegraphics[width=0.32\linewidth]{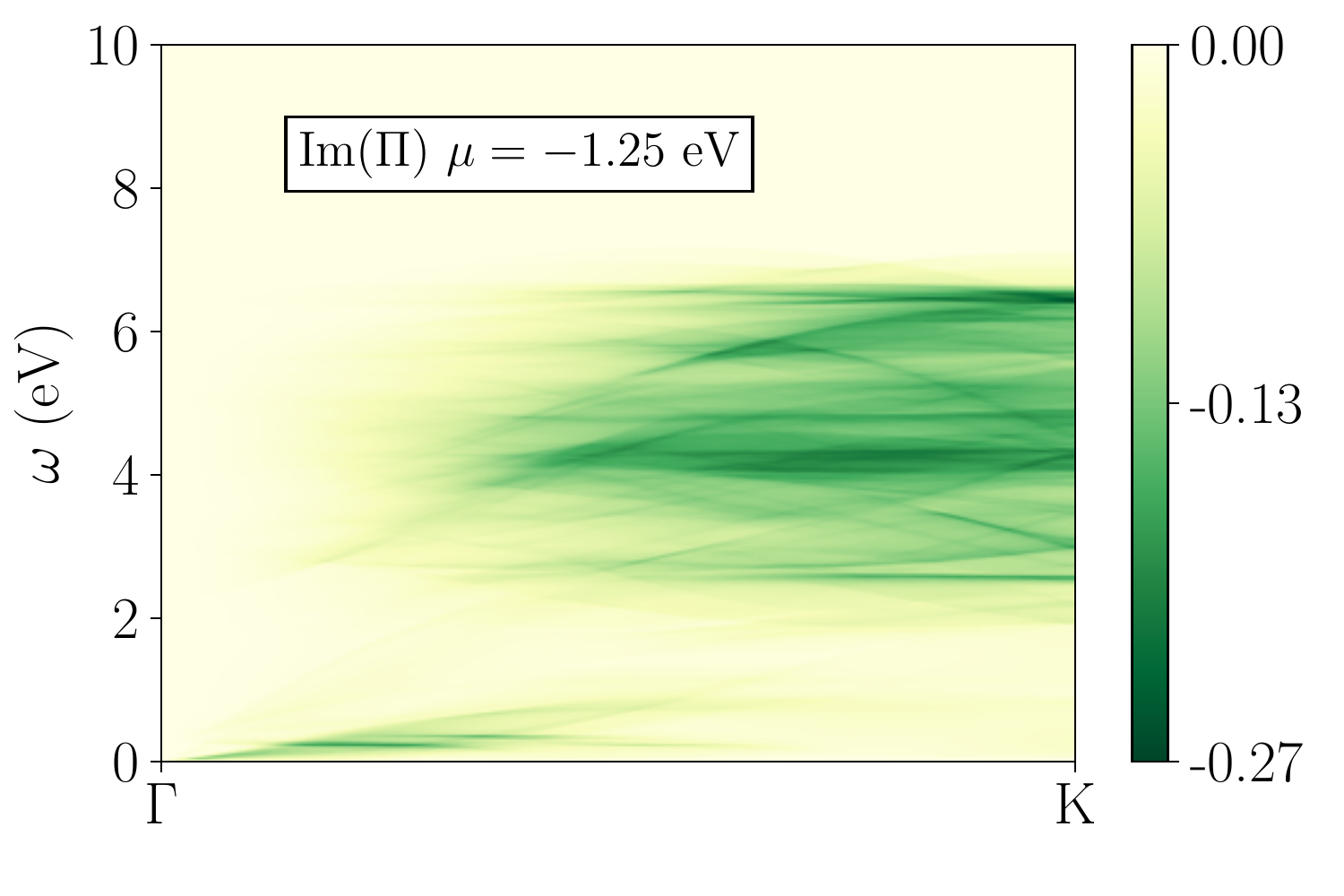}

    \caption{Electron-hole excitation spectrum of single layer antimonene. Imaginary part of the polarization function ${\rm Im}~\Pi^0({\bf q},\omega)$ for different values of the chemical potential $\mu$, with $\bf q$ along the $\Gamma$-K line in the BZ. The dotted black lines represent the $\bf q$ values depicted in the bottom images of Fig.~\ref{fig:dp_imag_lineplots}. }  
\label{fig:dp_imag_contour}
\end{figure*}

\begin{figure*}[ht!]
    \includegraphics[width=0.32\linewidth]{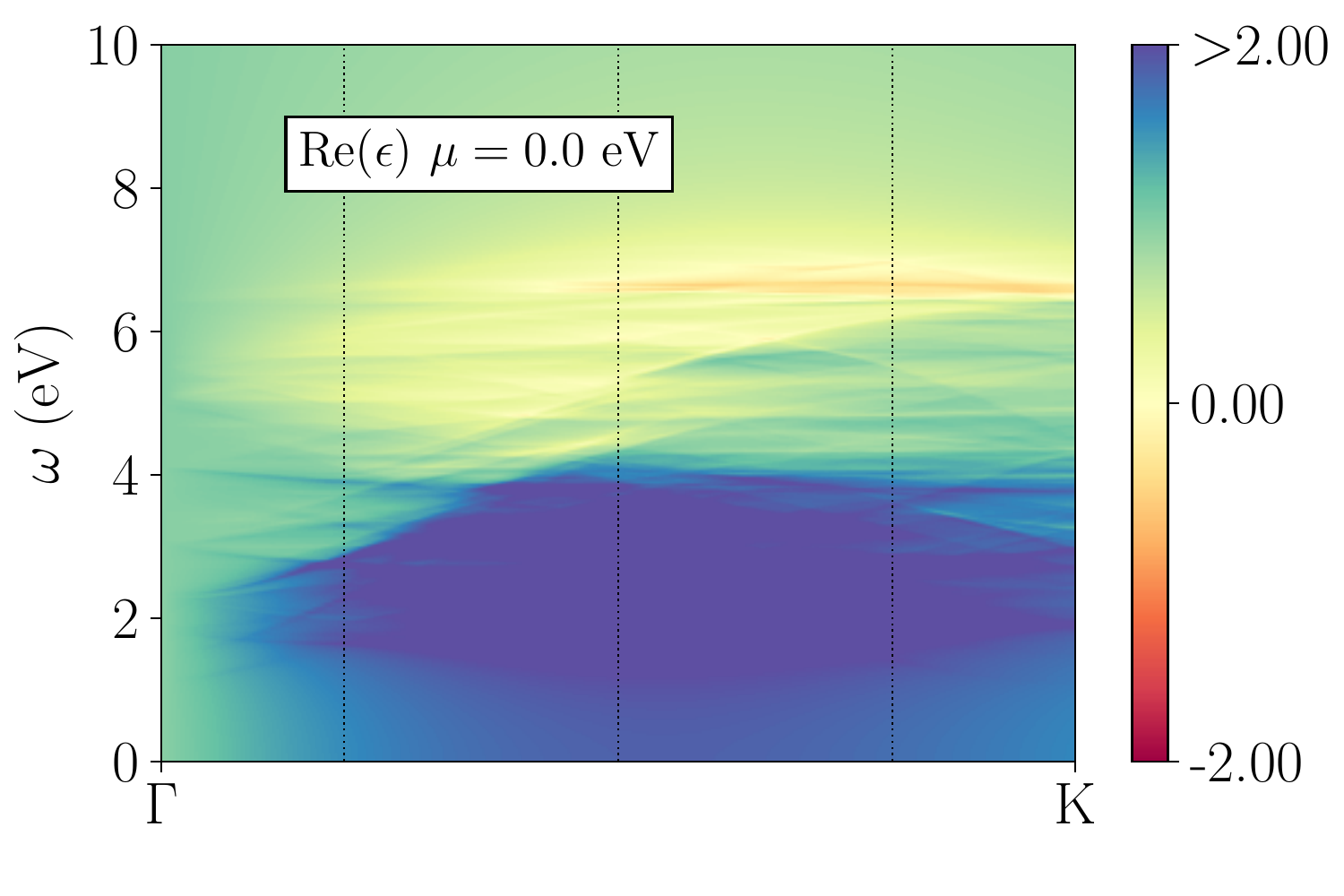}
    \includegraphics[width=0.32\linewidth]{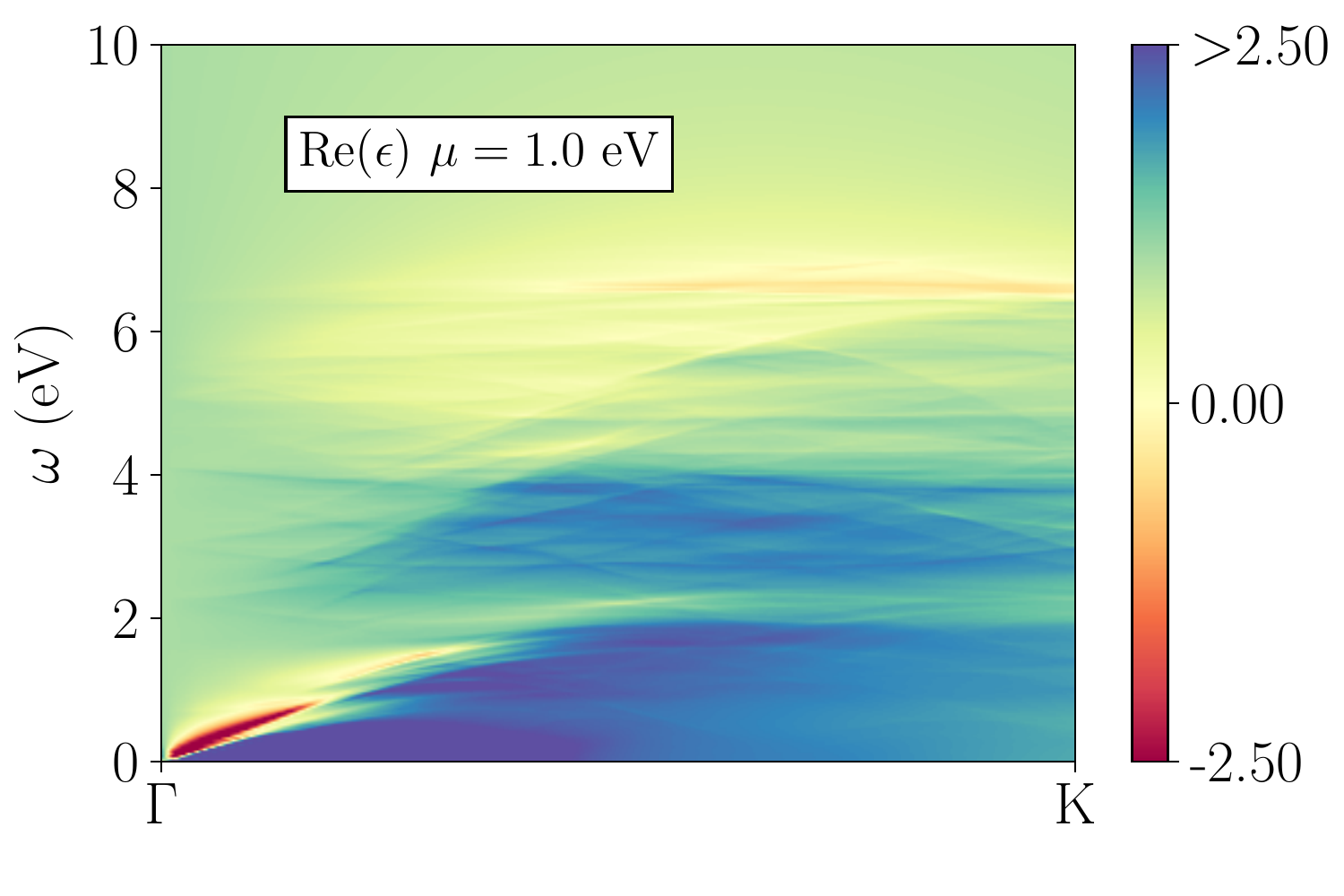}
    \includegraphics[width=0.32\linewidth]{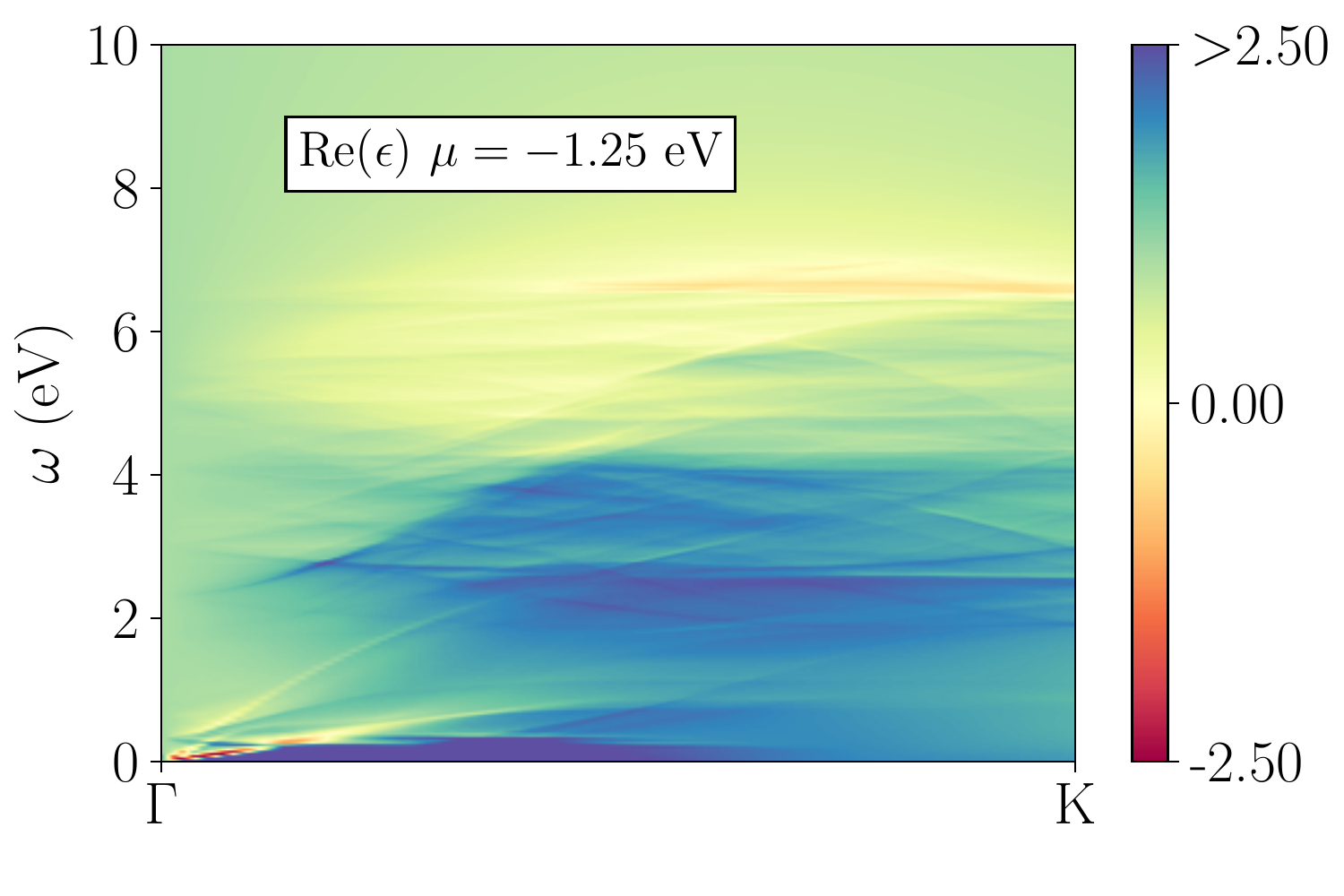}

     \includegraphics[width=0.32\linewidth]{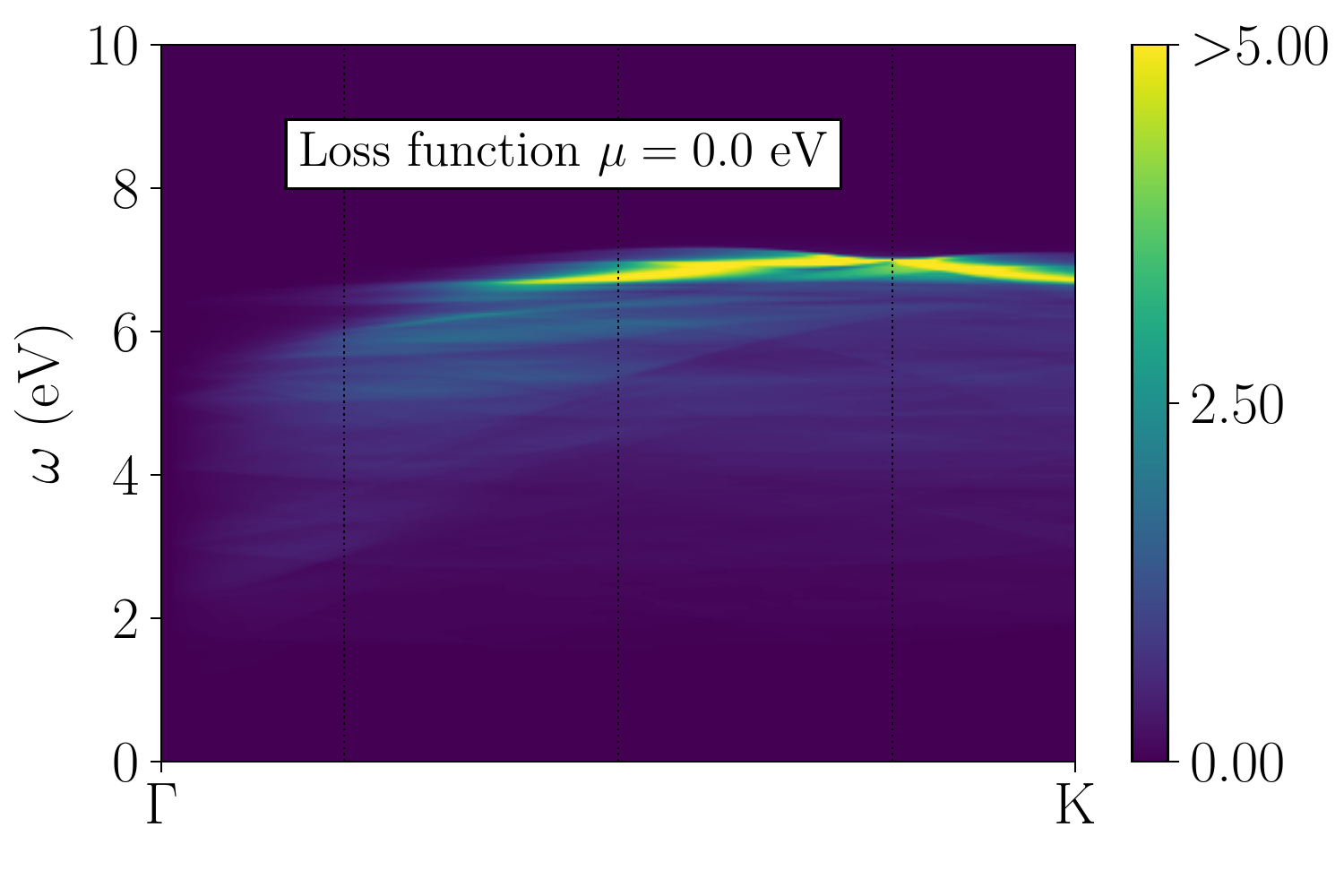}
    \includegraphics[width=0.32\linewidth]{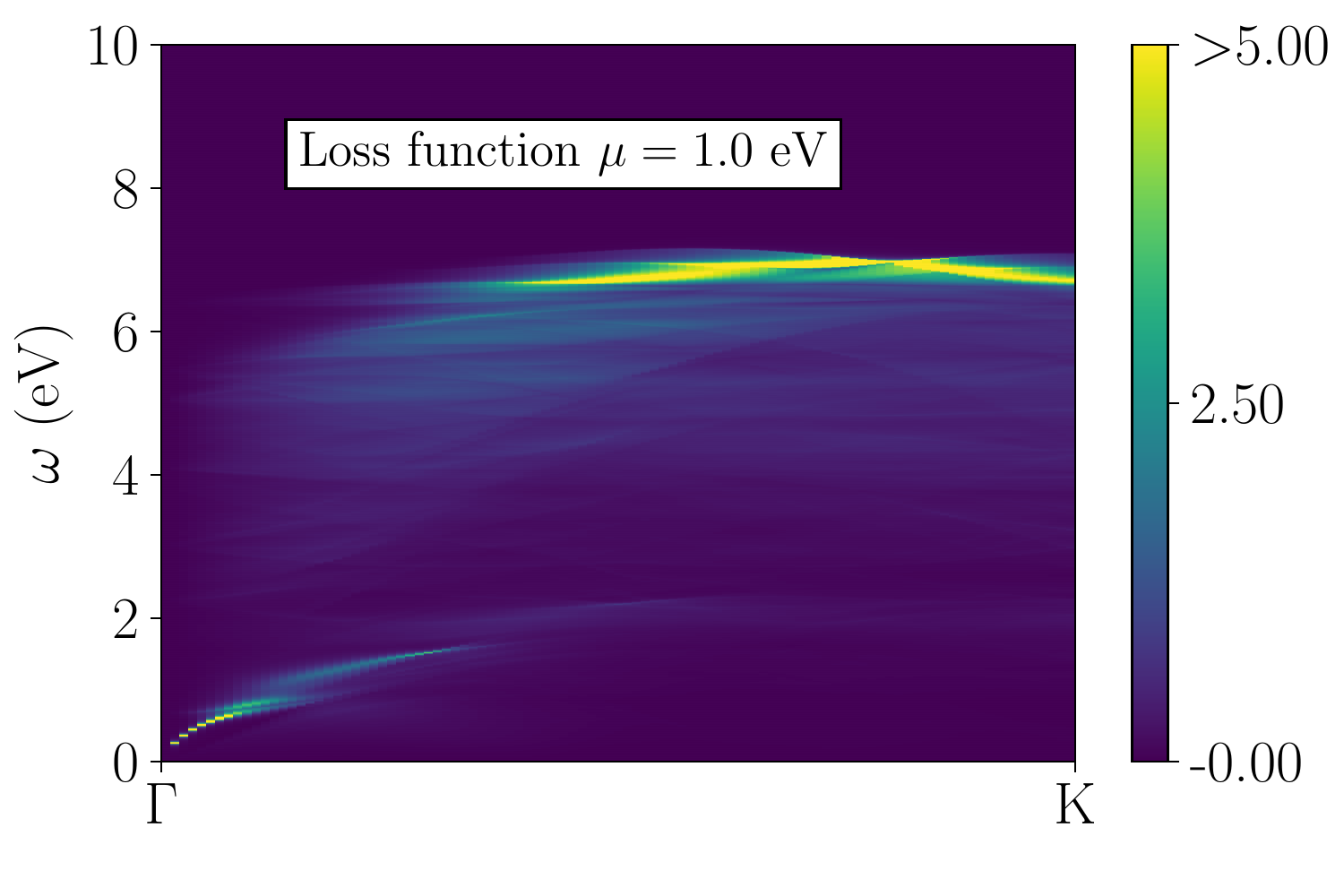}
    \includegraphics[width=0.32\linewidth]{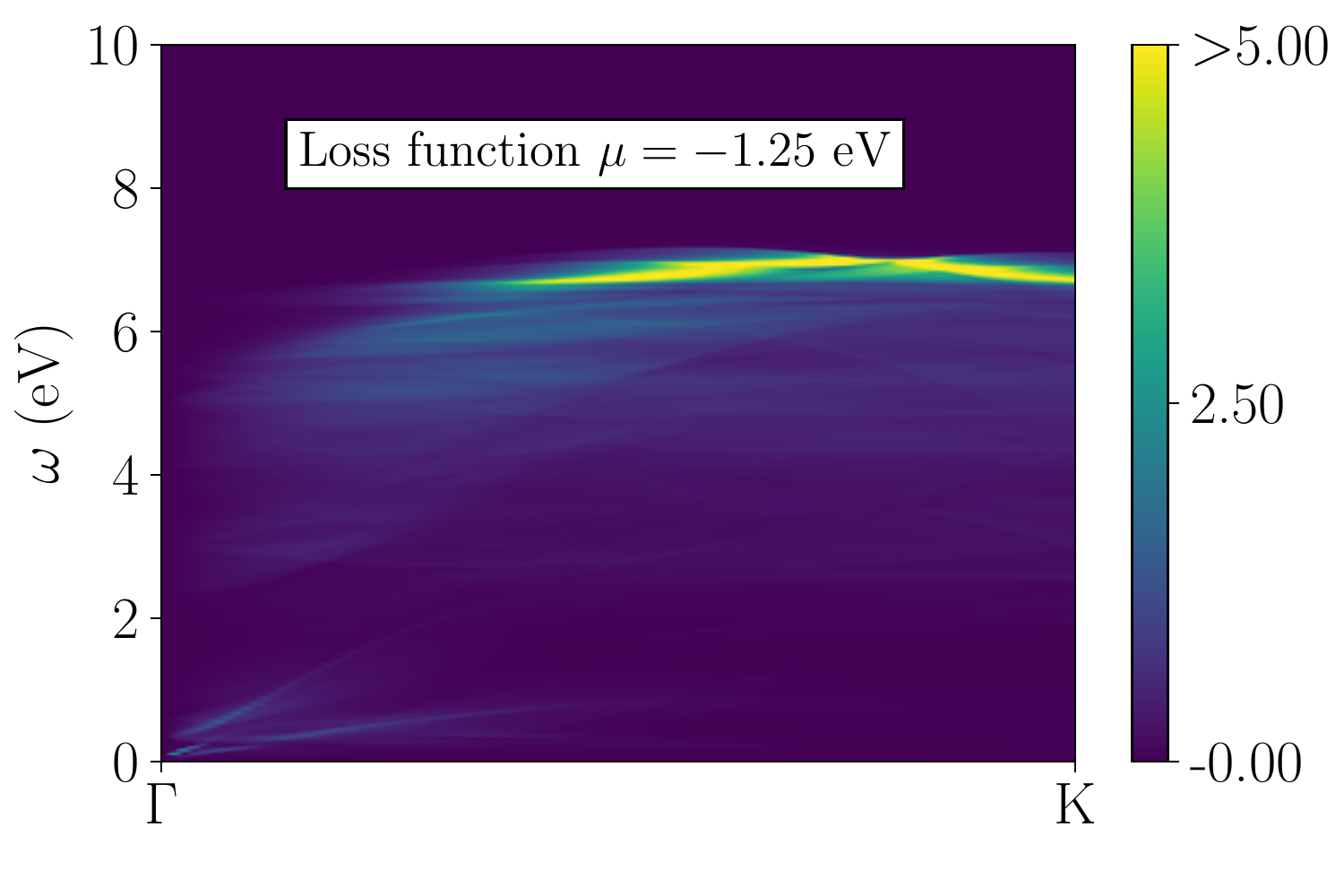}

    \caption{Real part of the dielectric function $\epsilon (\mathbf{q},\omega)$ (top panels) and the loss function ${\rm -Im}(1/\epsilon (\mathbf{q},\omega))$ (bottom panels), for different values of the chemical potential $\mu$, with $\bf q$ along $\Gamma$-K of the Brillouin zone. The black dotted lines indicate the values of $\bf q$ used in Fig. \ref{fig:ep_real_lineplots}. 
    }\label{fig:ep_real_contour}
\end{figure*}

To obtain more information about the collective modes, in this appendix we calculate the loss function, defined as $-{\rm Im}(1/\epsilon)$. Fig. \ref{fig:dp_imag_contour} shows the non-interacting electron-hole continuum of single-layer antimonene in the $\omega-{\bf q}$ plane, and Fig.~\ref{fig:ep_real_contour} shows the loss spectrum. Notice that, whereas in the dynamical polarization and in the dielectric function, the spectrum consists of many different peaks of similar amplitude, in the loss function there are some modes that clearly dominate. For these, both the imaginary part and the real part of the dielectric function tend to zero, giving rise to well defined modes in the loss spectrum, which could be directly observed with experimental measurements such as EELS.

\bibliography{Sb_bibliography,Bib_2D}

\clearpage

\end{document}